\documentstyle[epsfig]{elsart}

\def\be{\begin{equation}}
\def\ee{\end{equation}} 
\newcommand{\ber}{\begin{eqnarray}} 
\newcommand{\eer}{\end{eqnarray}}

\def\fm3{fm$^{-3}$} 
 
\begin{document} 
\begin{frontmatter} 
\title{The likelihood of GODs' existence: Improving the SN 1987a  
constraint on the size of large compact dimensions} 
\author[ntg,int]{Christoph Hanhart,\thanksref{hanhart}} 
\author[suny]{ Jos\'e A. Pons,\thanksref{pons}} 
\author[ntg,ou] {Daniel R. Phillips,\thanksref{phillips}} 
\author[int] {Sanjay Reddy,\thanksref{reddy}}  
\address[ntg]{Department of Physics, University of Washington, Box 351560,  
 Seattle, WA 98195-1560. } 
\address[int]{Institute for Nuclear Theory, University of Washington,  
Box 351550, Seattle, WA 98195-1550. } 
\address[ou]{Department of Physics and Astronomy, Ohio University, 
Athens, OH 45701.} 
\address[suny]{Department of Physics and Astronomy, SUNY at Stony Brook,  
Stony Brook, New York 11794-3800.} 
\thanks[hanhart]{E-mail: hanhart@phys.washington.edu} 
\thanks[phillips]{E-mail: phillips@phy.ohiou.edu} 
\thanks[pons]{E-mail: jpons@neutrino.ess.sunysb.edu} 
\thanks[reddy]{E-mail: reddy@phys.washington.edu} 
\maketitle
\begin{abstract} 
The existence of compact dimensions which are accessible only to gravity 
represents an intriguing possible solution to the hierarchy problem. At present 
the strongest constraint on the existence of such compact Gravity-Only 
Dimensions (GODs) comes from SN 1987a.  Here we report on the first 
self-consistent simulations of the early, neutrino-emitting phase of a 
proto-neutron star which include energy losses due to the coupling of the 
Kaluza-Klein modes of the graviton which arise in a world with GODs. We 
compare the neutrino signals from these simulations to that from SN 1987a and 
use a rigorous probabilistic analysis to derive improved bounds for the radii 
of such GODs. We find that the possibility that there are two compact extra 
dimensions with radii larger than 0.66 $\mu$m is excluded at the 95\% 
confidence level---as is the possibility that there are three compact extra 
dimensions larger than 0.8 nm. 
 
\end{abstract} 
\end{frontmatter} 
 
\section{Introduction} 
The possibility that there exist large, compact, space-like dimensions 
accessible only to gravity, but not to the standard model particles, 
has recently been suggested as a solution to the hierarchy 
problem~\cite{Arkani-Hamed:1998rs,Arkani-Hamed:1999nn}. If such 
compact spatial dimensions do exist then the fundamental Planck scale 
could be close to the electro-weak scale, with gravity becoming 
comparable in strength to other interactions at energies of order a 
few TeV.  At present, the strongest constraint on the existence of 
compact Gravity-Only Dimensions (GODs) such as these is derived from 
the SN 1987a neutrino signal.  The argument, put simply, is that if 
too much of the available energy is radiated to the GODs then the 
neutrino signal seen on terrestrial detectors would have differed 
markedly from that actually observed. This argument allowed 
Arkani-Hamed {\it et al.} (ADD) to place order-of-magnitude bounds on 
the radii of the GODs~\cite{Arkani-Hamed:1999nn}. Subsequently, an 
attempt to improve the analysis of Ref.~\cite{Arkani-Hamed:1999nn} was 
made by Cullen and Perelstein in Ref.~\cite{Cullen:1999hc}, who found 
$R < 0.3 \, \mu$m ($R<0.4$ nm) for 2 (3) compact extra 
dimensions. Further improvements to the calculation of the crucial 
Kaluza-Klein (KK) graviton emissivities were made in 
Ref.~\cite{Hanhart:2000er}. These calculations tended to confirm that 
the original estimate of Ref.~\cite{Arkani-Hamed:1999nn} was indeed 
within an order of magnitude of the correct number, although they 
resulted in bounds a factor of two or three weaker: $R<0.71 \, \mu$m 
($R<0.85$ nm).  However, as indicated in Ref.~\cite{Cullen:1999hc}, to 
obtain truly accurate bounds on the size of the GODs---and, by 
inference, the fundamental Planck scale---one would have to 
incorporate the contribution to the emissivities for the graviton 
emission to the extra dimensions into a numerical code for PNS 
evolution. In contrast, Refs.~\cite{Cullen:1999hc,Hanhart:2000er} both 
used the simple criterion suggested by Raffelt~\cite{Raffelt:1996}, 
which requires only that the emissivity (energy radiated per unit mass 
and per unit time) of some exotic particle be less than $10^{19}$ 
ergs/g/s. 
 
This bound is based on the time-scale of several seconds over which 
the handful of electron-type anti-neutrino events from SN 1987a were 
detected in Kamiokande~\cite{Hirata:1987hu} and 
IMB~\cite{Bionta:1987qt}.  These detections confirmed the standard 
scenario of core-collapse Supernovae: neutron stars are born in the 
aftermath of successful supernova explosions, as the stellar remnant 
becomes gravitationally decoupled from the expanding 
ejecta~\cite{Burrows:1986me,Burrows:1988ba,Keil:1995hw,Pons:1999mm,Pons:2000xf}. 
The ambient conditions in a newly born neutron star, also called a 
proto--neutron star (PNS), are so extreme---densities of order 
$10^{14}$ g/cm$^3$ and temperatures between 10 and 50 MeV---that the 
neutrino mean-free path is much smaller than the stellar 
radius. Consequently, the enormous gravitational binding energy gained 
during the stellar collapse ($\sim 2-3 \times 10^{53}$ ergs) is stored 
inside the PNS and slowly released by neutrino diffusion on a time 
scale of tens of seconds.  This picture meshes nicely with the long 
duration of SN 1987a's neutrino signal, since the neutrinos must have 
diffused, and not free-streamed, out of the PNS.  In this context, any 
weakly-interacting particle which couples to nucleons and {\it can} 
freely stream out of the star easily competes with neutrinos as a 
means of transporting energy away from the stellar interior.  Such an 
additional heat sink would accelerate the cooling and remove a 
significant fraction of the binding energy. The late-time neutrino 
signal, which is fueled by the heat stored in the core, is 
particularly sensitive to these effects, and it would be drastically 
altered if the couplings to exotic particles were strong enough. 
 
It is this physics that facilitates the use of the supernova neutrino 
signal to place bounds on the amount of energy lost as exotic 
radiation. Stringent bounds have been placed on the properties of a 
number of novel, weakly-interacting, light-particle species. 
Initially, these included limits on the neutrino magnetic moment, the 
mass of the axion, the strength of right-handed neutrino interactions 
and the sterile-neutrino production 
rates~\cite{Goldman:1988fg,Raffelt:1988yt,Turner:1988by,Lattimer:1988mf,Barbieri:1988nh,Burrows:1989ah}. 
Subsequently, several authors have employed Raffelt's bound on the 
total emissivity due to these exotic energy-loss mechanisms---or 
analogous bounds on the 
luminosity~\cite{Kolb:1996pa}---as a simple means of enforcing the 
constraint provided by the SN 1987a neutrino signal. However, since 
the emission rates are strong functions of temperature, the bounds on 
exotic-particle couplings which are deduced from such arguments will 
depend crucially on the fiducial value of temperature at which the 
exotic-particle emissivity is calculated. Not surprisingly, since the 
temperature of the newly-born neutron star varies with both position 
and time, the fiducial temperature is rather loosely defined, with 
values between 10 and 70 MeV appearing in the literature. Furthermore, 
exotic cooling mechanisms will lower the temperature inside the PNS 
core, and so the appropriate fiducial temperature is really a function 
of the emissivity of these processes.  The strong temperature 
dependence of these emissivities ultimately means that these ambiguities can 
easily change the resulting bound by more than an order of magnitude. 
 
Thus in order to establish more accurate bounds it is necessary to 
perform detailed simulations of the effect of exotic radiation on the 
neutrino signal. So far such simulations have only been carried 
out for the case of axion 
radiation~\cite{Burrows:1989ah,Keil:1997ju}. These simulations tend to 
confirm the picture outlined in the previous two paragraphs. Therefore 
in this work we report on the first self-consistent simulations of the 
PNS core which test the effect of GODs on the SN 1987a signal. In \S 2 
we present a summary of the early stages of the life of a neutron star 
and the effect of additional energy loses. In order to make these 
simulations as realistic as possible we use the most recent 
calculations of the KK-graviton emissivity~\cite{Hanhart:2000er}, 
which differ by roughly a factor of five from earlier, more schematic, 
calculations~\cite{Cullen:1999hc}. 
 
Another important drawback of former analyses is the lack of a well-established 
selection criteria to disregard a given model.  As emphasized in 
Ref.~\cite{Jegerlehner:1996kx}, the sparseness of the SN 1987a data calls for a 
Likelihood analysis. In \S 3 we outline the basics of the likelihood formalism 
which allows for marginalization over uncertain parameters, as well as the 
derivation of a bound with a clear probabilistic interpretation and 
unambiguously-defined confidence levels. Our results, and a discussion of them, 
are presented in 
\S 4. 
 
\section{PNS Evolution with Additional Sources of Energy Loss 
and SN 1987a signal} 
 
Core-collapse supernovae are incredibly rich physical systems, which include 
many complicated phenomena. Indeed there is still much debate about basic 
features like the details of the explosion 
mechanism~\cite{Janka:2000bt,Rampp:2000ws}.  What is, however, now 
well-established is that the birth and subsequent evolution of the PNS located 
in the inner region of the supernova provides the main source of neutrinos 
during the first several seconds after core bounce.  Nearly all of the binding 
energy gained during the collapse is trapped inside the PNS and this energy is 
radiated by neutrino diffusion over several tens of 
seconds~\cite{Burrows:1986me,Burrows:1988ba,Keil:1995hw,Pons:1999mm}.  
These time scales and 
the associated neutrino luminosities are affected by several pieces of physics, 
including the total mass of the PNS, and both the nuclear equation of state 
(EOS) and neutrino opacities at supra-nuclear density. None of these quantities 
can be determined in a model-independent approach, but earlier studies have 
explored the model sensitivities by using different EOSs and the associated 
self-consistently-calculated opacities~\cite{Pons:1999mm}.  These studies 
indicate that the critical physical parameter that determines neutrino 
luminosity in the first several seconds is the PNS total mass. The 
uncertainties associated with the properties of dense hadronic matter do not 
affect these luminosities greatly, provided that the opacities are calculated 
consistently with the EOS, because important feedbacks between the EOS and the 
opacities tend to reduce the differences.  The protoneutron star evolution code 
employed for this study has been described in detail in 
Refs.~\cite{Pons:1999mm,Pons:2000xf}.  Details regarding the neutrino opacities 
employed can be found in Ref.~\cite{Reddy:1998yr}.  In a recent paper the rates 
for the emissivities of KK-gravitons in dense matter were computed in a model 
independent way using low-energy theorems which relate the emissivities to the 
well measured nucleon-nucleon cross sections~\cite{Hanhart:2000er}. In a world 
with $n=2$ or $n=3$ compact GODs the KK-graviton emissivity of neutron-star 
matter can be fitted to within 5\% accuracy over a range of temperatures from 
15 to 30 MeV and a range of densities from $0.5 n_0$ to $3 n_0$, where 
$n_0=0.16$ fm$^{-3}$ is the nuclear saturation density. The result, which 
includes contributions from $nn$, $pp$, and $np$ collisions is: 
\begin{equation} 
\frac{dE}{dt} = a_n \left(\frac{n_B}{n_0}\right) 
\left(\frac{T}{10~{\rm MeV}}\right)^{p_n} \chi(X_n,X_p) 
\quad {\rm MeV/baryon/s} 
\label{em-gr} 
\end{equation} 
with $p_2=5.42$ , $a_2 = 5.1 \times 10^{4} R_2^2$, $p_3=6.5$, $a_3 = 1.4 \times 
10^{16} R_3^3$. Here, $n_B$ is the baryon number density, $T$ is the 
temperature, and the size of the extra dimensions, $R_n$, is given in 
mm~\footnote{In the non-degenerate limit Ref.~\cite{Hanhart:2000er} derived an 
analytic formula with $p_2=5.5$, $p_3=6.5$, and $a_2=5.6 \times 10^4 R_2^2$, 
$a_3=1.5 \times 10^{16} R_3^3$. One might perhaps be justified in using these 
formulae, but instead we have chosen to use more accurate fits to the full 
result of Ref.~\cite{Hanhart:2000er}.}. The emissivity depends on the  
composition of matter through the function  
\begin{equation} 
\chi(X_n,X_p) = X_n^2 + X_p^2 + 4 X_n X_p  
\left(\frac{T}{10~{\rm MeV}}\right)^{-0.44}, 
\end{equation} 
where $X_n$ is the neutron fraction, and $X_p=(1-X_n)$ is the proton 
fraction. The weak, residual, temperature dependence in the function 
$\chi$ arises because the $np$ cross section decreases as 
approximately $1/ \sqrt{E_{cm}}$, where $E_{cm}$ is the center of mass 
energy of the colliding nucleons, while the $nn$ cross section is 
roughly independent of energy, for the energies of interest.  Over the 
range of values of proton fraction and temperature encountered in the 
PNS the function $\chi$ changes by less than $10$\% from the case of 
pure neutron matter, when $\chi=1$. Thus, the total emissivity is a 
weak function of the proton fraction $X_p$, and in our simulations we 
assume, for simplicity, that $X_n=1$. 
 
In this paper we will put limits on permitted values of the coefficients $a_2$ 
and $a_3$, assuming the temperature-dependencies $T^{5.42}$ and $T^{6.5}$. 
Thus, ultimately our results are valid for the coupling to {\it any} exotic 
particle which results in these temperature-dependencies. By placing limits on 
the coefficients $a_n$, we are led, in this case, to bounds on $R_n$---the 
radii of the extra dimensions---but the simulations, philosophy, and likelihood 
formalism discussed here can be applied in a much broader context.  The count 
rate in the detectors is sensitive to the neutrino luminosities and 
spectrum. The procedure employed to obtain the count rates is described in 
detail in earlier work \cite{Pons:1999mm,Pons:2000xf}.  In addition, here we 
include the effects due to the neutron-proton mass difference on the 
anti-neutrino absorption cross sections and correct the fiducial mass of the 
IMB detector to 6 ktons \cite{Schramm:1990}. Both lead to better agreement 
between the two detectors. 
 
The equilibrium diffusion approximation employed in the numerical code  
provides a fair description of the total neutrino luminosity and its time 
structure. However, additional assumptions are required if an 
anti-neutrino spectrum is to be derived from this information. In 
particular, to obtain an average anti-neutrino temperature---denoted 
here by $T_{\bar{\nu}_e}$---we assume that the neutrinos leaving the 
protoneutron star are thermalized inside a neutrinosphere and have a 
Fermi-Dirac spectrum with zero chemical potential. The 
neutrinosphere was chosen to be at an optical depth of 2/3 in 
accordance with the usual definition. However, it must be borne in 
mind that the neutrino mean-free paths are energy dependent, and thus 
the concept of a neutrinosphere is only approximate. In the 
next section we will study the effect of those uncertainties related 
to this definition of the {\it neutrino temperature} by exploring the 
sensitivity of the count rates and the deduced bounds to changes in 
our prescription. 
\begin{figure}[t] 
\centerline{{\epsfxsize=3.5in \epsfbox{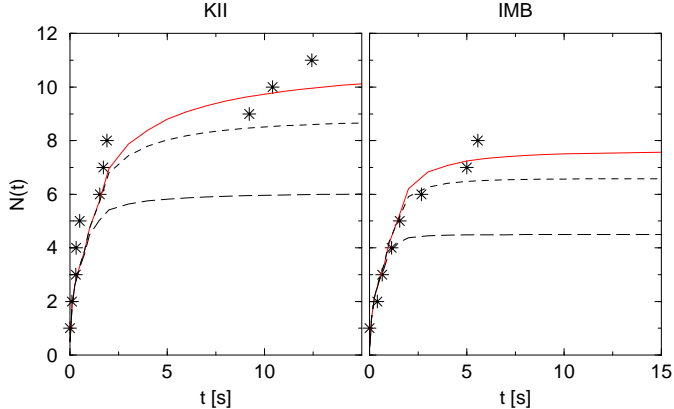}} } 
\caption{\it Comparison of the results of our simulations for a 
PNS with a baryonic mass of $M=1.6 ~M_\odot$ to the data from 
SN 1987a.  N(t) denotes the integrated number of counts. We  
show results from a model without graviton emission (solid line), as well as 
the results with graviton emission for $a_2 = 0.01$ MeV/baryon/s  
(short-dashed line) and $a_2 = 0.1$ MeV/baryon/s (long-dashed line).} 
\label{datacomp} 
\end{figure} 
In Fig.~\ref{datacomp} we show the number of accumulated counts as a function 
of time, $N(t)$, in Kamioka (left panel) and IMB (right panel). The results 
shown are for a baryonic mass of 1.6 M$_\odot$ (with a final gravitational mass 
of 1.46 M$_\odot$), and an anti-neutrino temperature defined as above. They 
include the case without KK-graviton emission, and two cases with radiation to 
the GODs included, namely: $a_2 = 0.01$ MeV/baryon/s and $a_2 = 0.1$ 
MeV/baryon/s.  As argued in the introduction, KK-gravitons steal energy from 
the core and thereby dramatically suppress the late-time neutrino 
emission. Note that the early-time neutrino emission is not strongly influenced 
by the existence of GODs because these neutrinos are emitted from the lower 
density and lower temperature regions of the star where the graviton emissivity 
is small. However, after about 2--4 seconds, neutrino losses provoke a strong 
compression and heating of the star 
\cite{Burrows:1986me,Keil:1995hw,Pons:1999mm}, which activates the KK-graviton 
emissivity. Consequently, the total number of neutrino counts, which clearly 
depends on the total amount of energy radiated in neutrinos, drops as $a_2$ is 
increased. 
 
This total number of counts also has a significant dependence on the PNS 
mass. Indeed, the mass of the PNS is generally the most important factor in the 
neutrino signal~\cite{Pons:1999mm}. However, it is worth noting that, if a 
sufficiently important additional sink of energy is included in the PNS 
simulation, the time-structure of the signal actually becomes less sensitive to 
the precise value of the PNS mass. This happens because more massive stars 
attain higher temperatures, which in turn results in enhanced graviton 
emission. Consequently, much of the additional binding energy reservoir in a 
more massive star is carried out by exotic particles, and the resultant 
neutrino luminosity is not dissimilar to that from PNSs of lower mass.  This 
was already noticed in Ref.~\cite{Keil:1997ju} for the axion case. 
 
\section{From simulations to bounds: the desirability of a likelihood 
analysis} 
 
Nevertheless, this ``mass-independence'' only arises for large values  
of the couplings $a_2$ and $a_3$. More typically we face the problem  
of trying to extract bounds for these couplings from the neutrino data  
in the face of significant uncertainties in the PNS mass $M$ and the  
anti-neutrino temperature $T_{\bar{\nu}_e}$. Therefore a key  
question is how to perform a consistent analysis of the results  
and derive a bound on the coupling of the exotic particles. Studies  
similar to ours for the case of axion emission 
\cite{Burrows:1989ah,Keil:1997ju}, based 
their investigations on either the time required to accumulate $90\%$ of  
the counts ($\Delta t(90\%)$) or the total number of counts detected 
$N^{\rm tot}$.  
 
Not surprisingly, both $N^{\rm tot}$ and $\Delta t(90\%)$ decrease with 
increasing $R_2$. The variation in the total number of counts is moderate and 
very sensitive to the variation of all three parameters studied: $a_2$, 
$T_{\bar{\nu}_e}$ and $M$. Thus, it makes more sense to study an observable 
that is relatively insensitive to $T_{\bar{\nu}_e}$ and $M$.   
$\Delta t(90 \%)$ is such a quantity.   
However, it is still difficult to move from the values 
of $\Delta t (90\%)$ to a confidence level for a bound on $R_2$. One might 
argue that a simple criterion is to disregard those radii $R_2$ which lead to 
$\Delta t(90\%)$ lower than one half of the value for the  
$a_n = 0$ case. Using this criterion we find that:  
\be 
R_2 \ < \ (0.4 \ ... \ 0.9) \ \mu\mbox{m} 
\nonumber 
\ee 
for PNS masses in the range $M \ = \ (1.5 \ .. .\ 2.0) \ M_{\odot}$.   
However, there is no guarantee that the result for $a_n=0$ is the 
``correct'' answer.  The performance of a model that badly overestimates the 
number of counts in the $a_n=0$ case might well be improved by the presence of 
GODs. So, if we adopt an alternative approach and disregard those values of 
$R_2$ which lead to a time scale that is less than one half of the value 
found in the SN 1987a data we find:  
\be 
R_2 \ < \ (0.2 \ ... \ 0.4) \ \mu\mbox{m} \ . 
\nonumber 
\ee 
These, however, are rather arbitrary 
criteria. There is really no statistical basis~\cite{Raffelt:1990yz} for 
judging what constitutes a successful reproduction of the experimental data, 
and thus no way to define confidence levels. 
 
In light of these issues we seek an analysis tool that will allow us to average 
over parameters which are not well-known (such as the mass of the PNS and 
$T_{\bar{\nu}_e}$) and also give a clear criterion that allows us to decide 
what values of the parameter space are excluded by the data. To this end we now 
pursue a ``likelihood'' analysis.  The goal of this analysis is the probability 
that the coupling $a_n$ is smaller than some specified value $a_n^0$, given the 
data, and a certain set of model assumptions.  We denote this probability by 
$\mbox{prob}(a_n \le a_n^0|\{ data \},I)$, where the $I$ indicates the set of 
physical assumptions which underly the simulations discussed in the previous 
section. Note that the assumptions denoted by $I$ do {\it not} include 
information about $M$ or $T_{\bar{\nu}_e}$.  Our goal is to relate 
$\mbox{prob}(a_n \le a_n^0|\{ data\} ,I)$ to a function we can calculate 
directly from our model.  That function is the Likelihood function  
${\cal L}_D (\{ data\} |a_n,M,T_{\bar{\nu}_e},I)$,  
which is the the probability that the 
data actually taken at a detector $D$ (KII or IMB) arises 
given a set of model assumptions ($I$) and certain specific values for the key 
quantities $M$, $T_{\bar{\nu}_e}$, and $a_n$~\footnote{We could, of course, 
have allowed for explicit variation of other parameters of the PNS simulation, 
and not simply included them in $I$, as we do when we calculate  
${\cal L}_D (\{ data\} |a_n,M,T_{\bar{\nu}_e},I)$.  
The formalism can easily be extended in this 
way if it becomes clear that some other physical parameter affects the neutrino 
signal strongly.}. Assuming that the data obeys Poisson statistics, the 
likelihood for the neutrino signal in a single detector can then be expressed 
in terms of the model prediction for the count rates at the 
times where an event happened and the total number of events actually 
seen~\cite{Loredo:1989} 
\begin{equation} 
{\cal L}_D(\{ data\} |a_n,M,T_{\bar{\nu}_e},I) =  
\left[ \prod_{i=1}^{N^{\rm tot}}\frac{dN_D(t_i^D)}{dt}\Delta t 
\right] e^{-N_D} \ ,  
\label{probdens} 
\end{equation} 
when $N_{D}$ is the total number of observed neutrino arrivals, and 
$t_1^D,t_2^D,\ldots$ are the times at which neutrinos actually arrived in the 
detector $D$.  $N_D(t)$ is the total number of neutrinos that the model with 
assumptions $I$, PNS mass $M$, anti-neutrino temperature $T_{\bar{\nu}_e}$, and 
exotic-coupling $a_n$ predicts will have arrived in the detector $D$ up until 
the time $t$. $\Delta t$ can be any interval small enough that the probability 
of detecting more than one count in any one bin can be taken to be 
negligible. Ultimately it will be absorbed into an overall normalization 
constant. Note that in order to get the full answer for the likelihood  
${\cal L}$ the product of the likelihoods for KII and IMB must be taken  
(i.e., ${\cal L} = {\cal L}_{KII} \times {\cal L}_{IMB} $).

The likelihood function provides us a with quantitative tool to 
compare models of PNS evolution.  Since we treat not only the coupling 
to extra dimensions, $a_n$, but also $M$ and $T_{{\bar \nu}_e}$, as 
parameters, the likelihood function is a function in a 
three-dimensional space. This function has a minimum at a baryonic 
mass of $M=1.5 M_\odot$, $a_2=0$ and $T_{\bar{\nu}_e}=1.1 T_\nu^o$.  
$T^o_\nu$ denotes the 
``reference anti-neutrino temperature'' defined above using the optical-depth 
prescription. We can now assess all other models by looking at the log 
of the ratio of their likelihood of a particular model to this ``most 
likely'' model: 
\begin{equation} 
q(a_n,M,T_{\bar{\nu_e}})=-\log \left(\frac{{\cal L}(\{ data\} 
|a_n,M,T_{\bar{\nu}_e},I)}{{\cal L}(\{ data\} 
|0,1.5 M_\odot,1.1 T_\nu^o,I)}\right) 
\label{eq:q} 
\end{equation} 
The function $q$ is then a function in this same three-dimensional 
space.  
 
\begin{figure}[t] 
\centerline{\epsfxsize=5.5in \epsfbox{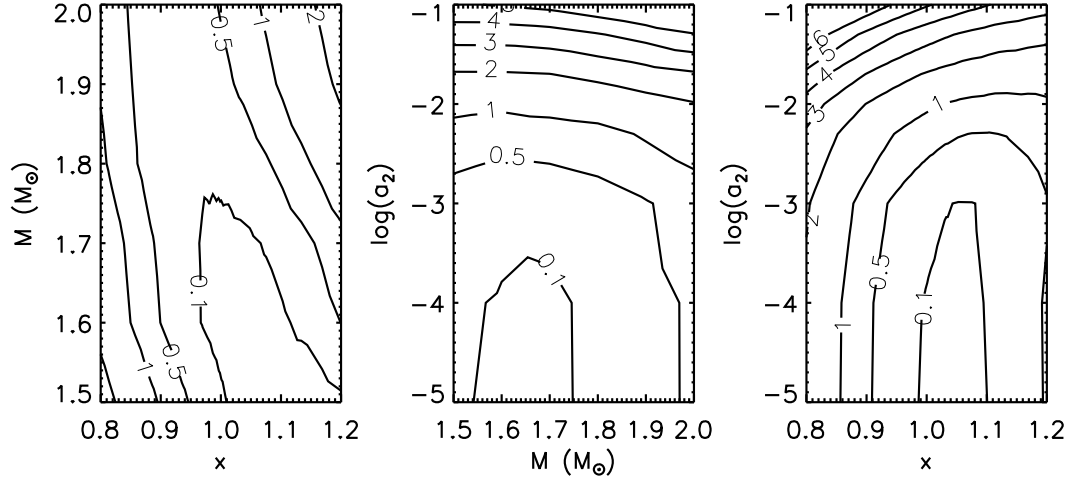}} 
\caption{\it Contour levels of the function $q$ defined in Eq. (\ref{eq:q}):  
$(x-M)$ left panel; $(M-\log(a_2))$ middle panel; $(x-\log(a_2))$ right panel.} 
\label{cont}  
\end{figure}

For the case of two GODs we discuss first the situation where these 
dimensions have zero radius, and so $a_2=0$.  Contours of $q$ in the 
resulting two-dimensional $x-M$ plane are shown in the left panel of 
Fig.~(\ref{cont}). Here the horizontal axis is the temperature 
normalized to the reference anti-neutrino temperature, $x=T_{{\bar 
\nu}_e}/T_\nu^o$, which is varied to explore the sensitivity of the 
results to deviations of the neutrino temperature from that obtained 
using the optical depth prescription.  We see that although $M=1.5 
M_\odot$, $T=1.1 T_\nu^o$ is indeed a minimum of the function $q$ it 
is a rather shallow minimum: varying the neutron-star mass and the 
anti-neutrino temperature over the range considered here does not 
reduce the likelihood greatly. Ultimately this reflects the weakness 
of the constraint that the SN 1987a data provides for these 
parameters. 
 
The situation is rather different as we move away from $a_2=0$. In the 
middle panel we display similar likelihood contours in the $M-\log(a_2)$ 
plane for the case $T_{{\bar \nu}_e}=T^o_\nu$, while the right panel 
shows contours in the $x-\log(a_2)$ plane for the baryonic mass  
$M=1.6 M_\odot$. (The 
pattern of contours shown is typical, and does not change significantly if 
different values of $T_{{\bar \nu}_e}$ and $M$ are chosen.)  These 
panels show that the likelihood function decreases rapidly for $a_2 
\ge 10^{-2}$---regardless of the values of the poorly-known parameters 
$M$ and $T_{{\bar \nu}_e}$. Such large values of $a_2$ are essentially 
two orders of magnitude less likely than the ``most likely'' 
model. This is a contrast to smaller values of $a_2$, where the 
differences in likelihood are comparable to those seen in the 
$M-T_{{\bar \nu}_e}$ plane. Thus statements about the most likely 
value of $a_2$ in this smaller-$a_2$ regime will be sensitive to the 
ill-constrained information on these neutron-star 
parameters. Consequently we will not make any such statements 
here. However, the two rightmost panels give us confidence in our 
ability to derive a {\it bound} on $a_2$---as opposed to a most-likely 
value---since it is clear that certain values of this coupling can be 
well-excluded, completely independent of details of the PNS modeling. 
 
With the likelihood function already in hand, the derivation of such a 
bound is quite a simple matter. A natural ansatz for the 
probability of $a_n$ to be smaller than some specified $a_n^0$ is  
\ber 
\mbox{prob}(a_n \le a_n^0|\{ data \} ,I) = N 
\int_{0}^{a_n^0}da\int_{0}^{\infty} dM \, g(M) \int_0^\infty 
dT_{\bar{\nu}_e} \, f(T_{\bar{\nu}_e},M) \nonumber \\ \ \times \ 
{\cal L}(\{ data\} |a_n,M,T_{\bar{\nu}_e},I). 
\label{finalmain2} 
\eer  
In other words, we now integrate (or ``marginalize'') over all possible values 
of $M$ and $T_{\bar{\nu}_e}$, using appropriate weight functions.  These 
functions are $f(T_{\bar{\nu}_e},M)$, which is the probability for the  
neutrinosphere to be at a temperature  
$T_{\bar{\nu}_e}$ given a PNS mass, $M$, and 
$g(M)$, which is the probability for that PNS mass to occur. The overall 
constant $N$ is defined by the normalization condition $\mbox{prob}(a_n \ge 
0|\{ data \} ,I) = 1$ . 
 
A rigorous derivation of this expression can be given by employing Bayesian 
statistics~\cite{Sivia:1996}.  In the course of this derivation the weight 
functions $f$ and $g$ acquire the aforementioned probabilistic  
interpretations.  Note that 
the structure of Eq.~(\ref{finalmain2}) is easily generalized: for every 
parameter not well determined from other sources one must sum over the 
parameter space with the appropriate relative weight.  In our case we will 
assume that the mass of the PNS lies within a certain range 
$[M_{min},M_{max}]$, but that within that range all values are equally 
probable. i.e., we write $g(M) \propto \theta(M-M_{min}) \theta(M_{max}-M)$. 
Meanwhile, the central value of the anti-neutrino temperature is calculated 
according to the optical-depth prescription, and $T_{{\bar \nu}_e}$ is then 
assumed o be Gaussian distributed with a 10\% width, independent of time.  It 
might be argued that a potential systematic error exists in the choice of 
anti-neutrino temperature, and thus a different weight function should be used. 
To check this we also used a function $f$ which was constant over a range of 
temperatures that differed by $\pm$ 20\% from the central value. This altered 
the bound obtained by less than 10\%. 
\section{Improved bounds and discussion} 
 
Equation (\ref{finalmain2}) may now be used to find bounds on $a_2$ 
and $a_3$. Defining the bound at the 95\% confidence level to be that 
value of $a_n^0$ which obeys $\mbox{prob}(a_n \le a_n^0|\{ data \},I) 
= 0.95$, we find that, for the $n=2$ case 
\begin{equation} 
a_2^0=0.022 \ \mbox{MeV/baryon/s and thus }R_2 \le 0.66 \, \mu\mbox{m.} 
\label{eq:R2} 
\end{equation} 
A similar analysis for the $n=3$ case differs in its details, but yields 
the same overall picture. The bound (again, at the 95\% confidence level) 
is: 
\begin{equation} 
a_3^0=0.0077 \ \mbox{MeV/baryon/s and thus }R_3 \le 8\times 
10^{-4} \mu\mbox{m.} 
\label{eq:R3} 
\end{equation} 
  These are within 10\% of the bounds derived in Ref.~\cite{Hanhart:2000er},  
where the emissivity (\ref{em-gr}) was derived and the Raffelt criterion  
applied using $n_B=n_0$ and a fiducial temperature chosen rather arbitrarily  
to be 30 MeV.  The good agreement between this work and  
Ref.~\cite{Hanhart:2000er} might be thought to suggest  that simple criteria
based on energetics are an entirely satisfactory.  However, analyses of our
simulations suggest that the  actual fiducial temperature in the $a_2=0$ case
was 35--40 MeV, and  the density was as large as 4$n_0$ for the most massive
stars. Application of  the same criterion at these temperatures and densities
could have resulted in   a bound on $a_2$ that was too stringent by a factor of
ten.   In contrast, the fiducial temperature in our simulations with $a_2>0.01$
was   generically less than 20 MeV, which leads to a  bound on $a_2$ that is
too weak by an order of magnitude. The advantage of our  analysis of the SN
1987a data is that it obviates any need to make a subjective  judgment about
the value of the ``real'' fiducial temperature.    Such an analysis confirms
that, for $n=2$ and $n=3$, the supernova  bound on the size of extra dimensions
found from the neutrino signal  of SN 1987a is significantly more stringent
than any collider bound  which will be obtained in the foreseeable future. It
is also, roughly, a factor of hundred more stringent than the bound derived by
means of recent tests of sub millimeter gravity \cite{Hoyle:2000cv} for the
case of two extra dimensions. We regard this astrophysical information as a
robust bound, since our 95\% confidence  level is fairly insensitive to
uncertainties in the neutron-star mass  and the anti-neutrino
temperature~\footnote{Arguments from 
cosmology~\cite{Hall:1999mk,Fairbairn:2001ct} may provide tighter  constraints
on $R_2$ and $R_3$, but there is much we presently do not  understand about
cosmology in the presence of GODs.}.  These bounds  are specific to scenarios
where the extra dimensions are flat or  weakly warped~\cite{Fox:2000mt}. They
come about because of the  existence of a large number of low-lying KK modes
with energy less the  characteristic temperature of the supernova core, and
thus do not  apply to extra-dimensional models where such low-lying KK modes
are  absent \cite{Randall:1999ee}.

We have not studied the extent to which changes in the neutrino
mean-free path or the equation of state of strongly-interacting matter
at high density influence our results. Recent work on the influence of
many-body correlations on the neutrino-transport mean-free path
indicates that the mean-field value for this quantity may be a factor
of two or three too small~\cite{Reddy:1999hb}.  However, as discussed
in Ref.~\cite{Keil:1997ju} in the axion-radiation case, increases in
the mean-free path actually strengthen the bound on exotic-particle
couplings, since the shorter neutrino diffusion times that result make
it harder to reconcile the late-time neutrino signal with the presence
of exotic radiation mechanisms. Consequently we view the bound derived
here as one that is conservative in regard to its assumptions about
neutrino mean-free paths.

A potentially more significant physics uncertainty is our lack of knowledge 
about the specific heat of matter at high density. Changes in this function 
might lead to a large effect on the bounds for $R_2$ and $R_3$, since they will 
produce different interior temperatures for the PNS, and the KK-graviton 
emissivity is a strong function of temperature. However, any changes in the 
PNS's interior temperature will also affect the neutrino diffusion time since 
neutrino mean-free paths are strongly temperature dependent. More information 
on the neutrino spectrum, neutron-star mass and interior physics of the PNS 
will undoubtedly shed further light on the problem. However, the likelihood 
function decreases rapidly as $a_2$ and $a_3$ are increased~(see 
Fig.~\ref{cont} for $a_2 \ge 0.01$), and so the bounds on these quantities may 
not be greatly affected by advances in our knowledge of the physics which 
governs the propagation of energy in the PNS interior.  
 
In contrast, the bound derived here {\it is} sensitive to the rate at which 
energy is radiated to the GODs, i.e. to the temperature-dependence of the 
graviton emissivity. In particular, many-body effects, such as the 
multiple-scattering suppression of bremsstrahlung reactions which occurs in a 
dense plasma and is known as the Landau-Pomeranchuk-Migdal effect, might 
strongly reduce the KK-gravistrahlung energy-loss rate. Indeed, earlier 
studies~\cite{Keil:1997ju} of the multiple-scattering suppression of axion 
radiation due to strong nucleon-spin fluctuations showed that the axion-mass 
bound was weakened by a factor of two when such effects were 
considered. Many-body effects might reasonably be expected to have a similar 
influence here---although gravitons couple to the stress-energy tensor, and 
not, as axions do, to the nucleon spin, and so they are sensitive to different 
correlations in the dense plasma.  All of these issues warrant further 
investigation and they will be addressed in future work.  In this work we have 
derived the bounds (\ref{eq:R2}) and (\ref{eq:R3}), which improve upon previous 
SN 1987a bounds on the size and scales of GODs in three important ways: 
 
\begin{enumerate} 
\item They represent the first self-consistent simulations of the 
early, neutrino-emitting phase of a proto-neutron star which also include 
energy losses due to the coupling of the Kaluza-Klein modes of the graviton 
have been employed to examine the SN 1987a neutrino signal in detail. 
 
\item They use the KK-graviton emissivities of Ref.~\cite{Hanhart:2000er}, 
thereby anchoring the nuclear physics of the emissivity for the $NN$ 
KK-gravistrahlung on solid ground. 
 
\item They are derived using a rigorous probabilistic analysis that 
facilitates an unambiguous criterion for the likelihood of certain values of 
the GODs' radii, and a consequent estimation of confidence levels. 
 
\end{enumerate} 
 
While these improvements do not alter the earlier estimated bounds
significantly they do serve to place them on a firm theoretical
basis. We also believe that in the event of a future galactic
supernova the formalism and ideas developed in this work will prove
useful in analyses of the large numbers of counts expected in
current-generation neutrino detectors, such as Super Kamiokande and
SNO.

{\it Acknowledgements:} We are grateful to M.~J.~Savage, W.~Haxton, 
J.~Lattimer and M.~Prakash  for 
useful discussions. J.~A.~P. and D.~R.~P. are grateful to the Institute for 
Nuclear Theory and to Iraj Afnan of Flinders University for their hospitality 
at various stages of this work. We thank the U.~S. Department of Energy for its 
support under DOE grants FG03-97ER4014, FG06-00ER41132 and FG02-87ER40317. 
C.~H.  acknowledges the support of the Alexander von Humboldt foundation. 
 
 
\end{document}